\documentclass[11pt,a4paper,oneside,reqno]{article}

\usepackage{amsmath,amssymb,graphicx} 
\usepackage{fullpage}
\usepackage{cite}
\usepackage{hyperref}
\usepackage{placeins}
\usepackage{subcaption}
\usepackage{authblk}
\usepackage[margin=1in]{geometry} 
\usepackage{enumerate}

\usepackage{amsmath}
\usepackage{nccmath}
\usepackage{float}
\usepackage[dvipsnames]{xcolor}

\usepackage{subcaption}
 \usepackage[utf8]{inputenc}
\usepackage[many]{tcolorbox}
\newtcolorbox{boxA}{
    fontupper = \bf,
    boxrule = 1.5pt,
    colframe = black 
}

\usepackage{hyperref}
\hypersetup{
colorlinks=true,
linkcolor=blue,
filecolor=magenta,
urlcolor=cyan,
citecolor=blue,
pdfpagemode=FullScreen
}
\usepackage{lscape}
\usepackage{graphicx}
\usepackage{soul}
 \usepackage[utf8]{inputenc}
\usepackage{placeins}
\usepackage{subcaption}
\usepackage{xcolor}
\title{Nonlinear Evolution of Anisotropic Matter Configurations under Higher-Order Curvature Corrections}
 \author{
 A. Zahra ${}^{(1,2)}$,\thanks{anam.zahra@vsb.cz}
 S. A. Mardan \thanks{syedalimardanazmi@yahoo.com}${}^{~(2,4)}$,
 Muhammad Bilal Riaz \thanks{muhammad.bilal.riaz@vsb.cz}${}^{~(1,3)}$,
 \thanks{bilalsehole@gmail.com}, Javlon Rayimbaev
${}^{(5,6,7,8)}$\thanks{javlonrayimbaev6@gmail.com; javlon@astrin.uz}, Inomjon Ibragimov ${}^{(9)}$ \thanks{i.ibragimov@kiut.uz}, Munisbek Akhmedov ${}^{(10)}$\thanks{munisbek95@urdu.uz},
 Erkaboy Davletov ${}^{(11)}$
\thanks{davletov\_erkaboy@mamunedu.uz}\\
 ${}^{1}$ IT4Innovations, VSB – Technical University of Ostrava, Ostrava, Czech Republic.\\
 ${}^{2}$ Department of Mathematics, University of the Management and Technology, Lahore, Pakistan\\
 ${}^{3}$ Applied Science Research Center, Applied Science Private University, Amman, Jordan.\\
 ${}^{4}$ Center for Theoretical Physics, Khazar University, 41 Mehseti Str., Baku, AZ1096, Azerbaijan\\
 ${}^{5}$Institute of Fundamental and Applied Research, National Research University TIIAME, Kori Niyoziy 39, Tashkent 100000, Uzbekistan\\
 ${}^{6}$ University of Tashkent for Applied Sciences, Str. Gavhar 1, Tashkent 100149, Uzbekistan\\
 ${}^{7}$  National University of Uzbekistan, Tashkent 100174, Uzbekistan\\
$^{8}$ Tashkent State Technical University, Tashkent 100095, Uzbekistan\\
${}^{9}$ Kimyo International University in Tashkent, Shota Rustaveli street 156, Tashkent 100121, Uzbekistan\\
${}^{10}$ Department of Technique, Urgench State University, Kh. Alimjan Str. 14, Urgench 221100, Uzbekistan\\
${}^{11}$ Mamun University, Bolkhovuz Street 2, Khiva, 220900, Uzbekistan}
 
\date{}
\begin{document}
\maketitle
\begin{abstract}
This study examines the dynamical evolution of self-gravitating systems in the presence of exotic matter within the framework of $f(R)$ gravity. Specifically, we have adopted the Starobinsky model $f(R) = R + \alpha R^2$, which incorporates higher-order curvature corrections to describe nonlinear gravitational behavior. The analysis focuses on the nonlinear spherical evolution of anisotropic matter configurations and explains how dark matter influences their physical characteristics. The presence of dark matter is found to significantly affect the radial and tangential pressure distributions, thereby altering the overall dynamics of the system. The model is employed for the compact object $ Her~X-1$ described by the generalized Tolman-Kuchowicz metric, demonstrating a singularity-free behavior of the physical parameters. The results reveal that increasing the parameter $n$ of the generalized Tolman-Kuchowicz metric leads to striking variations in the model characteristics, highlighting its essential role in governing internal structure and evolution of the compact object. The model remains physically viable under different testing criteria like energy conditions, hydrostatic equilibrium condition, adiabatic index, causality conditions, Herrera's Cracking condition and mass-radius relation presented in this work.
\end{abstract}

\section{Introduction}

Modified gravity theories have emerged as important extensions of Einstein's general relativity (GR). They aim to resolve several outstanding issues, including cosmic acceleration, galaxy rotation curves, and the origin of singularities in black holes and the early universe. These frameworks can explain cosmic acceleration without invoking dark energy (DE) and potentially reproduce effects of dark matter (DM) without requiring exotic particles by introducing additional geometric terms, scalar fields, or higher-order curvature corrections. One of the biggest problems in modern physics is the mysterious behavior of DE and DM. Recent findings from the Planck mission~\cite{1} indicate that the universe is composed of approximately 4.9\% ordinary (baryonic) matter, 26.8\% DM, and 68.3\% DE. DE is widely regarded as the primary driver behind the universe's accelerated expansion over cosmic time. Baryonic matter is ordinary matter made up of basic particles, protons and neutrons, which form atoms, stars, planets, and all visible structures in the universe. It interacts with light, gravitational, and electromagnetic forces. DM is a mysterious and invisible form of matter that neither emits, absorbs, nor reflects electromagnetic radiation, rendering it undetectable through conventional observational methods. Its existence is observed through gravitational effects on visible matter, such as the anomalous rotation curves of galaxies and the phenomenon of gravitational lensing. DE is a mysterious energy that is considered to be the cause of the universe's accelerating expansion and permeates all of space. Unlike ordinary matter, it does not cluster or interact with other forces except gravity. Modified gravity theories provide a testing ground for unifying gravity with quantum mechanics, offering more accurate models of compact astrophysical objects.

In \(f(R)\) gravity, the phenomena attributed to DM and DE can be reinterpreted as manifestations of spacetime geometry, where the Ricci scalar \( R \) is generalized to a function \( f(R) \). The inclusion of modified curvature terms in the gravitational field equations (FEs) provides a natural explanation for the universe's accelerated expansion. These modifications also explain the gravitational effects commonly attributed to DE and DM, instead of treating them as separate entities. To explain this phenomenon, it is essential to include additional elements in the matter of FEs, often known as DM and  DE (see references \cite{kamenshchik2001alternative}–\cite{Manzoor1} and references therein). Starobinsky \cite{Starobinsky} introduced this model and showed that quantum one-loop corrections to the Einstein equations yield non-singular, homogeneous, and isotropic cosmological solutions. The Starobinsky model stands as one of the most prominent and extensively studied inflationary frameworks in contemporary astrophysics  \cite{Percacci}-\cite{Aldabergenov}. It arises naturally within the framework of modified gravity, specifically in $f(R)$ gravity, and offers a geometric origin for early-universe inflation without introducing an explicit scalar field. These solutions imply that the universe may have originated in a highly symmetric de Sitter state, thereby avoiding the initial singularity. In this way, the interaction between topology and black hole thermodynamics in the rainbow of $f(R)$ gravity was examined by Sekhmani et al. \cite{Sekhmani}. The derivation of exact solutions to the FEs in extended gravity theories presents a significant theoretical challenge, primarily due to their inherent nonlinearity. Nonetheless, several physically viable solutions have been constructed, though many do not fully satisfy the stability and structural criteria required for realistic relativistic stellar configurations. 

The commonly employed techniques for modeling isotropic matter spheres are the Oppenheimer–Volkoff and Tolman methods. In contrast to the Oppenheimer–Volkoff approach, which depends on specific physical assumptions to close the system of equations, Tolman’s method employs an arbitrary selection for the metric potential \cite{Oppenheimer}. The equation of state (EoS) and other physical characteristics of stellar objects can be examined more easily due to this methodological freedom \cite{martin2004algorithmic,boonserm2005generating}. Recent studies have increasingly focused on stellar objects characterized by direction-dependent pressure profiles, driven by their intricate and compelling astrophysical properties \cite{sarkar2019compact}–\cite{tello2019anisotropic}. Maurya et al. \cite{maurya2020anisotropic} studied the properties of compact stars exhibiting pressure anisotropy within the Class I embedding scheme under extended gravity theories. In another contribution, Maurya et al. \cite{maurya2020non} utilized the extended geometric deformation approach to construct a class of stable spheroidal solutions, relevant to the internal structure of neutron stars. Singh et al. \cite{singh2020compact} investigated the stability of compact relativistic stars by adopting the Tolman metric as a foundational geometry and considering a nonlinear equation of state that incorporates contributions from dust and DE. Das et al. \cite{das2021modeling} focused on the slowly rotating compact object $4U~1820–30$, exploring the role of pressure anisotropy and employing the Bejger–Haensel empirical relation to estimate the moment of inertia. In recent years, a significant amount of research has developed exact analytical solutions to FEs for anisotropic, spherical stellar systems under a range of physically justified assumptions in both GR and alternative gravity frameworks \cite{zubair2021realistic}–\cite{lopes2019anisotropic}.

The generalized Tolman–Kuchowicz (GTK) metric offers a flexible framework for modeling an anisotropic star with non-singular behavior at the center. The GTK metric is important for constructing realistic star models that satisfy regularity and energy conditions. It enables the study of anisotropic matter in $f(R)$ gravity. Biswas et al. \cite{26} analyzed a gravitational model for strange anisotropic stars by utilizing Tolman-Kuchowicz (TK) metric potentials along with the MIT bag model EoS to characterize strange quark matter. The model undergoes several physical tests that confirm its stability and physical acceptability, with non-singular density and pressure profiles. Jasim et al.~\cite{27} constructed a static model for strange stars using the TK metric, incorporating a spatially dependent cosmological constant and the MIT bag model for strange-quark matter. The solution met all physical conditions for realistic stellar structure, demonstrating that higher values of the cosmological and bag constants lead to increasingly compact, ultra-dense stars. Bhar et al.~\cite{28} examined anisotropic compact stars within the framework of Einstein–Gauss–Bonnet gravity using TK spacetime. Their study focused on analyzing the influence of the coupling constant on the physical properties of these stellar configurations. Stability, equilibrium, and stiffness are analyzed and compared with GR, showing that higher values of the coupling constant increase stiffness but remain softer than GR. Acharya et al. \cite{29} studied the compact star PSR J0952–0607 by computing Einstein’s field equations (EFEs) using a GTK spacetime, which allows for a more flexible description of internal geometry. They adopted a quadratic EoS to model the connection between density and pressure, leading to stable and physically realistic stellar configurations. 

This study investigates the evolution of stellar structure of a star influenced by DM with the GTK metric, using the Starobinsky model for $f(R)$ gravity. A clearer motivation for adopting the Starobinsky $f(R)=R+\alpha R^2$ model together with GTK metric, and explicitly states the main objectives and theoretical significance of this study. This work is therefore useful because it provides an exact analytical framework for studying nonlinear, higher-order gravitational systems. It extends strong-field tests of modified gravity theories to realistic stellar configurations, bridging cosmological curvature corrections with astrophysical observables such as the mass-radius relation and stability profiles. This study contributes to the broader theoretical goal of explaining DE and DM phenomena as emergent geometric effects rather than external components. The main objectives of this study are 
(i) to formulate the modified FEs for anisotropic stellar configurations under the Starobinsky $f(R)$ gravity framework using the GTK metric, 
(ii) to study the influence of higher-order curvature terms and the GTK parameter $n$ on energy density, pressure anisotropy, regularity, energy, causality and equilibrium conditions, 
(iii) apply the framework to the observed neutron star $Her X-1$ and verify the consistency with empirical mass-radius constraints.
The structure of this paper is organized as follows.
Section~2 introduces the modified FEs in \( f(R) \) gravity for a self-gravitating stellar system. 
In Section~3, we focus on the Starobinsky model with the GTK metric and derive the corresponding modified FEs. 
Section~4 discusses the regularity conditions relevant to \( f(R) \) gravity. 
Section~5 addresses the exterior spacetime and establishes the boundary conditions applicable to the stellar configuration. 
Section~6 provides a detailed physical investigation of the system. 
Finally, Section~7 presents concluding remarks and a discussion of the results obtained.

\section{Derivation of modified field equations in $f(R)$ gravity }
In $f(R)$ gravity, the Einstein–Hilbert action can be expressed as
\begin{equation}\label{1}
    S_{f(R)} = \frac{1}{2\kappa} \int d^4 x \sqrt{-g} f(R) + S_M.
\end{equation}  
Here, the coupling constant is defined as \( \kappa = 8\pi G \) and the metric tensor is denoted by \( g \). The \( S_M \) represents the action corresponding to the matter fields, and the Ricci tensor is denoted by \( R \). By varying the Hilbert action with respect to the metric tensor \( g \) in Eq.~(\ref{1}), the \( f(R) \) gravity FEs are determined as follows
\begin{equation}\label{2}
    R_{\gamma \eta}f_R -\frac{1}{2}f(R)g_{\gamma \eta}+(g_{\gamma \eta}\Box-\nabla_\gamma\nabla_\eta)f_R=\kappa T_{\gamma \eta},
\end{equation}
\( \nabla_\gamma \) represents the covariant derivative, and \( \Box \equiv \nabla^\gamma \nabla_\gamma \) defines the d'Alembert operator. The term \( f_R \equiv \frac{df}{dR} \) represents the differentiation of \( f(R) \) with respect to \( R \).
 The Eq. (\ref{2}) can be rewritten as
\begin{equation}\label{3}
G_{\gamma \eta} = \frac{\kappa}{f_{R}}(T_{\gamma \eta}^{(D)}+T_{\gamma \eta})\equiv T_{\gamma \eta}^{(eff)},
\end{equation}
where the Einstein tensor is represented by $G_{\gamma \eta}$, $T_{\gamma \eta}$ is baryonic (ordinary) matter, and $T_{\gamma \eta}^{(D)}$ denotes the dark source (DM or DE). They can be expressed as

\begin{equation}\label{6}
 T_{\gamma \eta}=(\rho + p_{t})U_{\gamma}U_{\eta}-p_{t}g_{\gamma \eta}+(p_{r}-p_{t})V_{\gamma}V_{\eta}.
 \end{equation} 
The fluid's energy density is indicated here by (\( \rho \)).  The tangential pressure is represented by (\( p_t \)) and the radial pressure by (\( p_r \)).
 The four vectors \( U_\gamma \) and \( V_\gamma \) satisfy the orthogonality and normalization conditions as \( U^\gamma V_\gamma = 0 \), \( U_\gamma U^\gamma = 1 \), and \( V_\gamma V^\gamma = -1 \).
 
\begin{equation}\label{4}
T_{\gamma \eta}^{(D)}=\frac{1}{\kappa}\bigg\{\nabla _{\gamma}\nabla _{\eta}f_{R}-\Box f_{R}g_{\gamma \eta}+(f-Rf_{R})\frac{g_{\gamma \eta}}{2}\bigg\}.
\end{equation}
A static model the interior geometry of the compact object fluid, spherically symmetric line element, given as
\begin{equation}\label{5}
ds^2=e^{\nu(r)}dt^{2}-r^2(d\theta^2+sin^2\theta d\phi^2)-e^{\lambda(r)}dr^{2}.
\end{equation}

Using Eqs.~(\ref{3}) and (\ref{5}) the EFEs in  \( f(R) \) gravity are
\begin{eqnarray}\nonumber
\rho^{eff} &=&\frac{e^{-\lambda}}{2r^{2}}\Big(-2\lambda 'rf_{R}+2f_{R}-2f_{R}e^{\lambda}-2r^{2}f''_{R}+\lambda 'r^{2}f'_{R}-4rf'_{R}\\&+&r^{2}e^{\lambda}Rf_{R}-r^{2}fe^{\lambda}\Big),\label{7}\\
p_{r}^{eff}&=&\frac{e^{-\lambda}}{2r^{2}}\Big(2e^{\lambda}f_{R}-2f_{R}-r^{2}f_{R}e^{\lambda}R+r^{2}e^{\lambda}f+2f_{R}r\nu '+4rf'_{R}\nonumber\\&+&\nu 'r^{2}f'_{R}\Big),\label{8}\\
p_{t}^{eff}&=&\frac{e^{-\lambda}}{4r}\Big(-2f_{R}r\nu ''-\nu'^{2}rf_{R}-\nu '\lambda 'rf_{R}+2f_{R}\lambda '-2f_{R}\nu '+4f'_{R} \nonumber\\
&+&2\nu 'f'_{R}r+4rf''_{R}-2\lambda 'rf''_{R}+2e^{\lambda}rf-2e^{\lambda}rRf_{R}\Big),\label{9}
\end{eqnarray}
and the Ricci scalar is 
\begin{equation}\label{10}
    R=\frac{e^{-\lambda}}{2r^2}\Bigg(2r^2\nu''+r^2\nu'^2+4r\nu'-r^2\nu'\lambda'-4r\lambda'-4e^{\lambda}+4\Bigg).
\end{equation}
Here, 'prime' represents the derivative with respect to $r$.

\section{Stellar configurations in $f(R)$ gravity with the GTK metric and the Starobinsky model}
In the study of relativistic stellar stars such as $ Her~X-1$, the GTK model provides a valuable solution to EFEs by incorporating specific metric potentials that yield physically viable and singularity-free configurations. This model accounts for anisotropic pressure distributions and varying energy densities, making it suitable for describing the dense interiors of stars. When extended within the framework of modified gravity, particularly the Starobinsky model \( f(R) = R + \alpha R^2 \), the FEs gain higher-order curvature corrections that offer a geometric explanation for early-universe inflation and late-time acceleration. Integrating the Starobinsky model with the GTK metric allows for the exploration of stars under the influence of both nonlinear gravity effects and dark source dynamics, providing deeper insight into their structure, evolution, and potential deviations from standard GR. Recent research by Nazar and colleagues \cite{nazar2025possible,nazar2025exhibiting} has focused on the qualitative behavior of static, spherically symmetric compact objects exhibiting direction-dependent pressures. Their work involves analyzing various forms of the underlying spacetime within different gravitational theories. Aslam and Malik \cite{aslam2025impact} investigated the dynamical stability of compact stars with isotropic pressure profiles using the TK spacetime metric. Malik et al. \cite{malik2025modeling} adapted the same geometric configuration to construct models of compact stars consisting of conventional baryonic and exotic matter under a modified $f(R)$ gravity formulation.

The expression for the GTK metric is
\begin{equation}\label{gtk}
    e^\lambda=(1+ar^2+br^4)^n, \quad e^\nu= A^2e^{Br^2}.
\end{equation}
The parameter \( n \) is a positive real number that satisfies \( n \geq 1 \). The exponent \( n \) is included in the modified metric potential \( e^\nu \) in the GTK metric \cite{56}. When \( n = 1 \) \cite{44,45}, it reduces to the original TK form.
The constants \( A \), \( a \), and \( b \) possess units of \( \text{km}^{-2} \), \( \text{km}^{-2} \), and \( \text{km}^{-4} \), respectively, while \( B \) is taken to be dimensionless. 
The numerical values of the constants will be determined by ensuring smooth matching between the interior and exterior spacetimes. The chosen metric potentials are constructed to yield a regular and non-singular stellar configuration, the detailed analysis of which is presented in Section 5. Using these potentials, we derive the expressions for the $f(R)$ FEs relevant to the stellar structure under consideration.

 \begin{eqnarray}
\rho^{eff} &=&\frac{(1 + a r^2 + b r^4)^{-n}}{2r^2} \Bigg(
2 f_R- {(1 + a r^2 + b r^4)^{-n}} f_R r^2 
- 4 r f_R' - 2 r^2 f_R''
\nonumber \\&+& \frac{n r^2 (2 a r + 4 b r^3) f_R'}{1 + a r^2 + b r^4} 
- \frac{2 n r (2 a r + 4 b r^3) f_R}{1 + a r^2 + b r^4} - 2 {(1 + a r^2 + b r^4)^{-n}} f_R  
\nonumber \\&+& {(1 + a r^2 + b r^4)^{-n}} r^2 R f_R 
\Bigg),
\end{eqnarray}
\begin{eqnarray}
p_{r}^{eff}&=&- \frac{(1 + a r^2 + b r^4)^{-n}}{2 r^2} \Bigg(
2 f_R 
- 2 f_R (1 + a r^2 + b r^4)^n- 4 A r^2 f_R  - 4 r f_R'
\nonumber\\&-& f_R r^2 (1 + a r^2 + b r^4)^n 
+f_R R r^2 (1 + a r^2 + b r^4)^n  
- 2 A r^3 f_R'
\Bigg),
\end{eqnarray}
\begin{eqnarray}
p_{t}^{eff}&=& \frac{(1 + a r^2 + b r^4)^{-n}}{4r} \Bigg( 
 4 A r f_R -  \frac{2 An r^2(2 a r + 4 b r^3)}{1 + a r^2 + b r^4} f_R
+ 4 A r^2 f_R' 
 \nonumber \\&+& 4A^2r^3f_R  + 4 A rf_R 
-  \frac{2 rn(2 a r + 4 b r^3)}{1 + a r^2 + b r^4} f_R' 
- \frac{2n(2 a r + 4 b r^3)}{1 + a r^2 + b r^4} f_R
 \nonumber \\\nonumber&+& 2 r f_R (1 + a r^2 + b r^4)^n - 2 rR f_R (1 + a r^2 + b r^4)^{-n} 
+ 4 r f_R'' 
+ 4 f_R' 
\Bigg).\\
\end{eqnarray}
\subsection{Physical viability of matter components in \( f(R) \) gravity}
Relativistic compact stellar systems' stability and internal structure are significantly influenced by the $\rho$, $p_r$, and $p_t$. To be determined physically accurate, the quantities of a stellar model must behave appropriately throughout the stellar interior. Consequently, we carry out a detailed examination of these properties to assess the consistency and applicability of our proposed configuration.

\begin{figure}[ht!]
    \centering
    \includegraphics[width=0.60\linewidth]{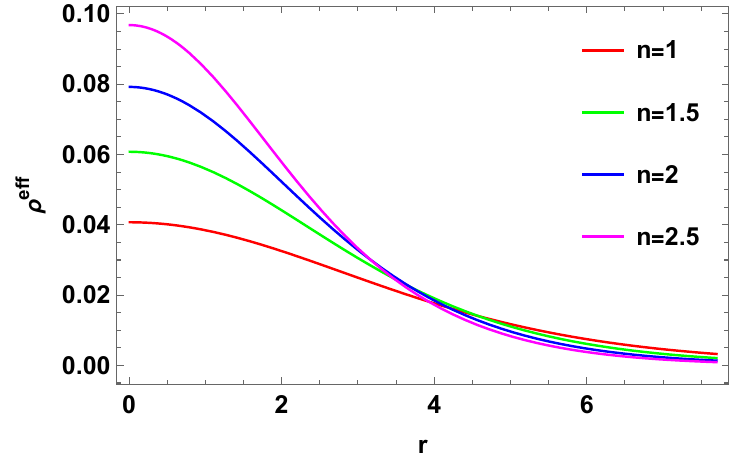}
    \caption{Graph of $\rho^{eff}~(km^{-2})$ with respect to radial variation $r~(km)$ for $ Her~ X-1$ through various GTK parametric values $n=1,~1.5,~2,~2.5$.}
    \label{fig:1}
\end{figure}

\begin{figure}[ht!]
    \centering
    \includegraphics[width=0.60\linewidth]{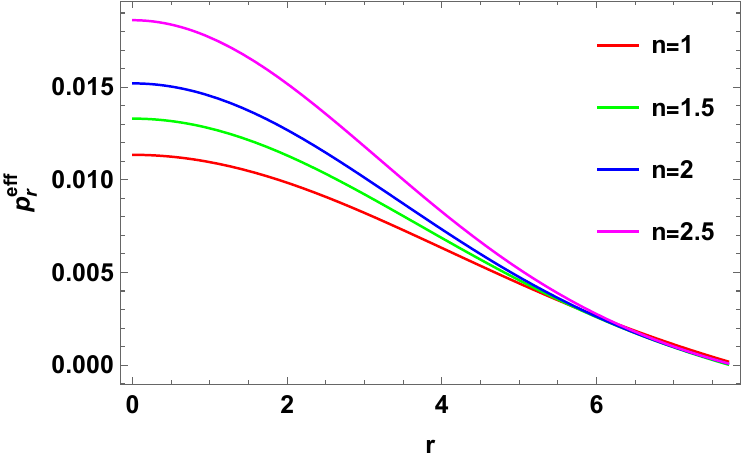}
    \caption{Graph of $p_r^{eff}~(km^{-2})$ with respect to radial variation $r~(km)$ for $ Her~ X-1$ through various GTK parametric values $n=1,~1.5,~2,~2.5$.}
    \label{fig:2}
\end{figure}

\begin{figure}[ht!]
    \centering
    \includegraphics[width=0.60\linewidth]{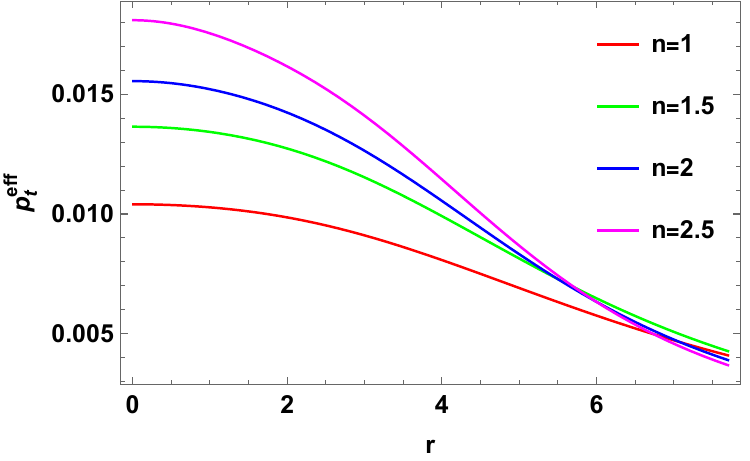}
    \caption{Graph of $p_t^{eff}~(km^{-2})$ with respect to radial variation $r~(km)$ for $ Her~ X-1$ through various GTK parametric values $n=1,~1.5,~2,~2.5$.}
    \label{fig:3}
\end{figure}

\begin{figure}[ht!]
    \centering
    \includegraphics[width=0.60\linewidth]{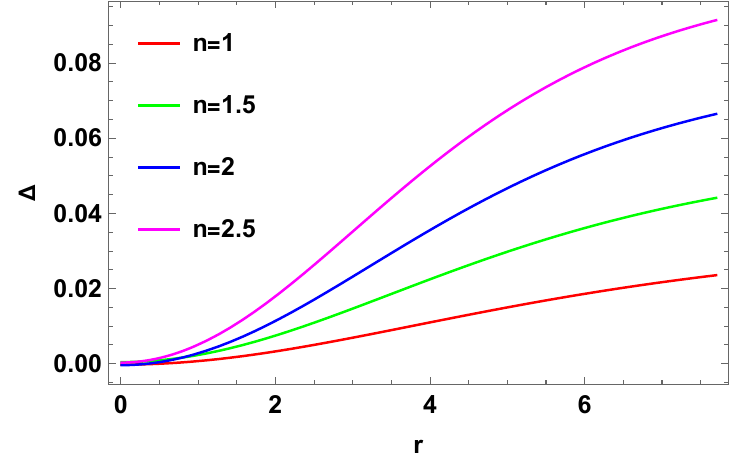}
    \caption{Graph of $\Delta~(km^{-2})$ with respect to radial variation $r~(km)$ for $ Her~ X-1$ through various GTK parametric values $n=1,~1.5,~2,~2.5$.}
    \label{fig:4}
\end{figure}

\begin{itemize}
    \item We analyze how the matter density evolves within a compact star for various values of the GTK parameter $n$, as illustrated in Fig. (\ref{fig:1}). This clearly demonstrates that the energy density remains finite and regular throughout the interior region. It reaches a maximum at the center of the star and then gradually drops towards the outer boundary, approaching a minimum as $r\rightarrow R$. This monotonically decreasing profile is characteristic of compact objects \cite{ruderman1972pulsars, glendenning2012compact,herzog2011three}. This behavior provides strong support for the suitability of the model in describing the $Her~ X-1$ stellar configuration.
    \item Variations in radial and transverse pressures for the distribution of matter $Her~ X-1$ are shown in Figs. (\ref{fig:2}) and (\ref{fig:3}). Both pressures maintain physically acceptable behavior—they are regular, positive, and finite throughout the star's interior. As expected, the radial pressure decreases from its central maximum and vanishes at the stellar boundary \cite{zel2014stars,misner1964relativistic}. In contrast, the transverse pressure remains finite and non-zero at the boundary, never becoming negative or divergent. This non-vanishing transverse pressure at the surface is indicative of the anisotropic and spheroidal characteristics typically associated with compact objects  \cite{shee2017compact,quevedo1989general,chifu2012gravitational}. 
    \item Fig. (\ref{fig:4}) illustrates the radial variation of the anisotropy $(\Delta)$ for $Her~X–1$, analyzed for various values of the GTK metric parameter $n$. The plot clearly shows that it starts from zero at the stellar center and increases monotonically, exhibiting a smooth and regular behavior throughout the star's interior. This trend is consistent with the physical requirement that the anisotropy vanishes at the center to maintain regularity. Notably, higher values of $n$ lead to greater magnitudes of $\Delta$, indicating that the degree of anisotropy strengthens with increasing $n$. For instance, the curve corresponding to shows the most pronounced growth, reaching a maximum anisotropy near the stellar surface.  Moreover, the fact that $\Delta>0$ throughout implies that the $p_t$ exceeds the $p_r$, generating a repulsive anisotropic force that can support more massive and compact stellar structures. This characteristic is particularly important for modeling strange-quark stars, where pressure anisotropy contributes to the stability and viability of ultra-dense matter configurations \cite{deb,astashenok1,hossein}.
\end{itemize}

\section{Regularity conditions in $f(R)$ gravity theory}
In the context of star modeled in 
$f(R)$ gravity, regularity conditions are essential to ensure that the physical and geometric quantities describing the system are well-behaved, finite, smooth, and physically meaningful at the center of the star and throughout the interior. Below are the standard regularity conditions typically required for spherically symmetric stellar models in $f(R)$ gravity \cite{zahra2025investigating}.

\begin{enumerate}
    \item The solution must have no geometric or physical singularities and maintain a finite, positive density and central pressure in order to be considered physically acceptable.  The metric potentials (\( e^\lambda \)) and (\( e^\nu \)) must be positive and non-zero throughout to satisfy \( e^\lambda(0) = 1 \) and \( (e^\nu)' = 0 \) at the center.
 
    \item At the core of the stellar star, all physical quantities such as  (\(\rho^{eff} \)), radial pressure (\(p_r^{eff} \)), and tangential pressure (\(p_t^{eff} \)) should be non-negative and finite. This ensures a physically realistic and stable stellar configuration.
    
    \item To ensure that anisotropy $ (\Delta=0)$ is in the center, $p_r^{eff}$ and $p_t^{eff}$ must be equal in the center (\( r = 0 \)), that is, \( p_r^{eff}(0) = p_t^{eff}(0) \).
    
    \item At the star's boundary the $p_r^{eff}$ must be zero, which means that \( p_r(R_{\Sigma}) = 0 \).
    
    \item Throughout the interior (\( 0 \leq r \leq R_{\Sigma} \)), the derivatives of $\rho^{eff}$, $p_r^{eff}$ and $p_t^{eff}$ must be negative as $\frac{d\rho^{eff}}{dr} \leq 0, \quad \frac{dp_r^{eff}}{dr} \leq 0, \quad \frac{dp_t^{eff}}{dr} \leq 0.$
    
    \item The radial $(v_r^2)$ and tangential  $(v_t^2)$ sound speeds must meet the causality conditions, which states that $0 \leq v_r^2 = \frac{dp_r^{eff}}{d\rho} \leq 1 \quad 0 \leq v_t^2 = \frac{dp_t^{eff}}{d\rho} \leq 1$.

    \item The adiabatic index \( (\Gamma_r) \) must exceed the critical value \(\frac{4}{3} \) throughout the stellar interior to ensure stability.

    \item All energy conditions ensure that the matter inside the star behaves in a physically realistic way, with non-negative energy density and pressure components.  
Their satisfaction confirms that the star's structure is stable, gravitationally attractive, and free from exotic or non-physical matter.
\end{enumerate}
\begin{figure}[ht!]
    \centering
    \includegraphics[width=0.60\linewidth]{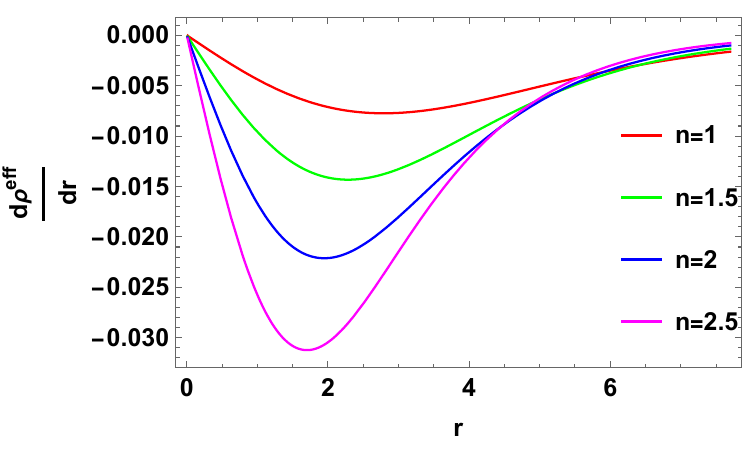}
    \caption{Graph of $\frac{d\rho^{eff}}{dr}~(km^{-3})$ with respect to radial variation $r~(km)$ for $ Her~ X-1$ through various GTK metric values $n=1,~1.5,~2,~ric5$.}
    \label{fig:5}
\end{figure}
\begin{figure}[ht!]
    \centering
    \includegraphics[width=0.60\linewidth]{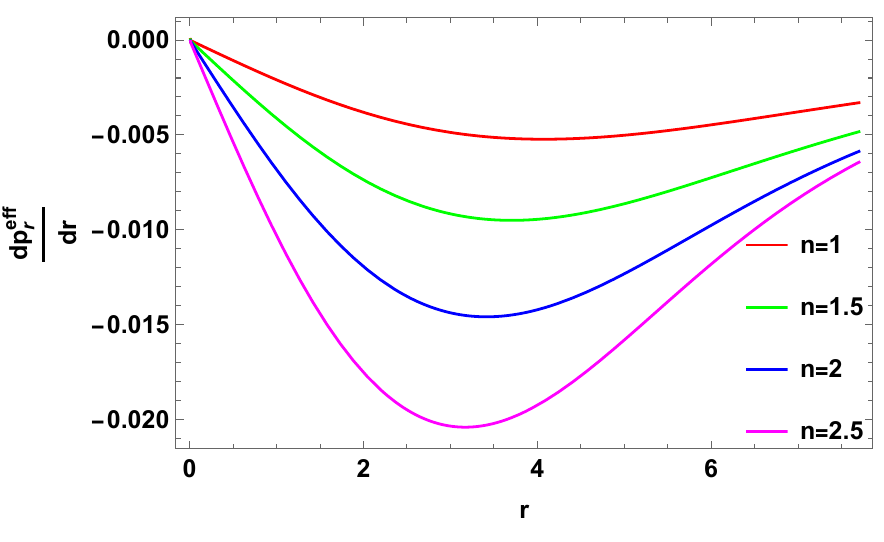}
    \caption{Graph of $\frac{dp_r^{eff}}{dr}~(km^{-3})$ with respect to radial variation $r~(km)$ for $ Her~ X-1$ through various GTK metric parametric values $n=1,~1.5,~2,~2.5$.}
    \label{fig:6}
\end{figure}
\begin{figure}[ht!]
    \centering
    \includegraphics[width=0.60\linewidth]{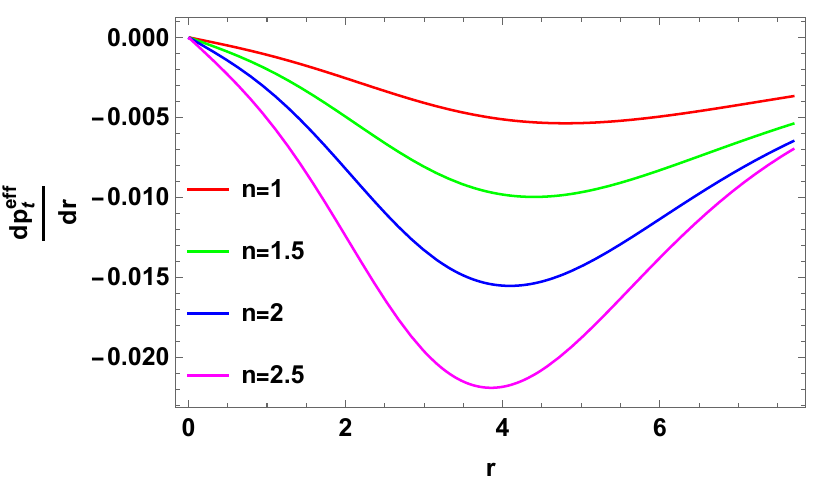}
    \caption{Graph of $\frac{dp_t^{eff}}{dr}~(km^{-3})$ with respect to radial variation $r~(km)$ for $ Her~ X-1$ through various GTK  parametric values $n=1,~1.5,~2,~2.5$.}
    \label{fig:7}
\end{figure}
\FloatBarrier
\section{The exterior spacetime and boundary conditions}
The boundary conditions are applied at the boundary \( r = R_{\Sigma} \) of the star to determine the unknown constants \( a \), \( A \), \( b \), and \( B \) in the model.  
 To guarantee physical continuity in the spacetime geometry, the interior solution described by the GTK metric must seamlessly connect with the exterior Schwarzschild solution at this boundary \cite{schwarzschild1999gravitational}.
\begin{equation}\label{19}
    ds^2 =g(r) \, dt^2 - \frac{1}{g(r)} \, dr^2 - r^2 \left( d\theta^2 + \sin^2\theta \, d\phi^2 \right),
\end{equation}
where \( M \) represents the star's total mass and $g(r)=\Big(1-\frac{2M}{r}\Big)$. At \( r = R_{\Sigma} \), ensuring continuity across the boundary requires that the metric components \( e^{\nu} \) and \( e^{\lambda} \), as well as the radial derivative of \( e^{\nu} \), match smoothly between the interior and exterior solutions~\cite{darmois1927equations,israel1966singular,chu2022generalized}, such that

\begin{equation}\label{12}
   1-\frac{2M}{R_{\Sigma}}=B^2e^{AR^2_{\Sigma}},
\end{equation}
\begin{equation}\label{13}
  \Big(1-\frac{2M}{R_{\Sigma}}\Big)^{-1}=(1+aR^2_{\Sigma}+bR^4_{\Sigma})^n,
\end{equation}
\begin{equation}\label{14}
 \frac{M}{R_{\Sigma}}=AR^2_{\Sigma}B^2e^{AR^2_{\Sigma}}.
\end{equation}
At the boundary the $p_{r\Sigma}$ must vanish, which provides
\begin{equation}
    p_{r}(R_{\Sigma})=0.
\end{equation}
The constants $A$, $B$ and $b$ can now be evaluated using Eqs. (\ref{12}-\ref{14}) as
\begin{equation}\label{15}
A = \left( e^{-\frac{A R^2_{\Sigma}}{2}} \right) \left(1 - \frac{2M}{R_{\Sigma}} \right)^{\frac{1}{2}},
\end{equation}
\begin{equation}\label{16}
B = \frac{M}{R^3_{\Sigma} \left(1 - \frac{2M}{R_{\Sigma}} \right)},
\end{equation}

\begin{equation}\label{18}
b = \frac{U^* - 1 - a R^2_{\Sigma}}{R^4_{\Sigma}}.
\end{equation}
Here, $U^*=\left(1 - \frac{2M}{R_{\Sigma}} \right)^{-\frac{1}{n}}$. By using Eqs.~(\ref{15}–\ref{18}), we can compute the constants  \( b \), \( A \), and \( B \).  Abubekerov et al.~\cite{abubekerov} examined the mass of the X-ray binary system $Her X-1$ using established physical methods. They reported the estimation of stellar mass \( m_x = 1.8M_\odot \)  based on radial velocity curve analyses. Such discrepancies are often attributed to X-ray heating effects in $Her X-1$. Numerous studies adopt these values of mass \( M \) and radius \( R \) for modeling various compact stars~\cite{85}-\cite{87}. Leahy and Mendelsohn \cite{leahy2025geometry} studied the $Her~X-1$ and mentioned the maximum mass from $1.52M_\odot$--$1.69M_\odot$. The mass-radius relation shown in Fig. (\ref{fig:14}) is consistent with the observational data presented in the literature. The model parameters for $Her X-1$, corresponding to a representative radius \( R \simeq 7.7\,\text{km} \)  are computed as follows:\\
\( A = 0.002498\,\text{km}^{-2} \), \( a = 0.02\,\text{km}^{-2} \), \( \alpha = 0.5 \), and \( B = 0.81561 \), where \( \alpha \) and \( B \) are dimensionless constants.
 Additionally, the values of the parameter \( b \) for various values of \( n \) are given by the following ordered pairs
$(n, b) = (1,~0.00059027\,\text {km}^{-4}),\ (1.5,~0.00055971\,\text{km}^{-4}), \\(2,~0.00054540\,\text{km}^{-4}),(2.5,~0.000537104\,\text{km}^{-4})$.

\section{Physical analysis of Starobinsky model with generalized Tolman--Kuchowicz metric}
The physical analysis of the Starobinsky model with the GTK metric provides insights into the behavior of star configurations under higher-order curvature corrections. It ensures regularity, stability (via the TOV equation), the equation of state parameter, adiabatic index, causality conditions, the Herrera cracking technique, and realistic energy conditions in $f(R)$ gravity frameworks.
\subsection{Energy conditions}
For a compact star model to be deemed physically realistic, the energy conditions must hold both throughout its interior and on its surface \cite{kolassis1988energy,hawking2023large,wald2010general,brassel2021inhomogeneous,brassel2021higher}.
The energy conditions relevant to this study are summarized in the following and illustrated in Figs.~ (8a-8e):

\begin{itemize}
\item {Dominant energy condition (DEC)}: ~\( \rho^{eff} - p_t^{eff} \geq 0 \),~\( \rho^{eff} - p_r^{eff} \geq 0 \), \\\( \rho^{eff} \geq 0 \)
 \item Strong energy condition (SEC): \( \rho^{eff} + p_r^{eff} \geq 0 \),~\( \rho^{eff}+ p_t^{eff}\geq 0 \),~\( \rho^{eff} + 3p_r^{eff} + 2p_t^{eff} \geq 0 \)
\item Weak energy conditions  (WEC): \( \rho^{eff}+ p_t^{eff}\geq 0 \),~\( \rho^{eff} + p_r^{eff} \geq 0 \),\\\( \rho^{eff}\geq 0 \)
\item {Null energy conditions (NEC):} \( \rho^{eff}+ p_t^{eff}\geq 0 \),~\( \rho^{eff}+ p_r^{eff}\geq 0 \)
\end{itemize}

\begin{figure}[!htbp]  
\centering
\subfloat[]{\includegraphics[width=70mm]{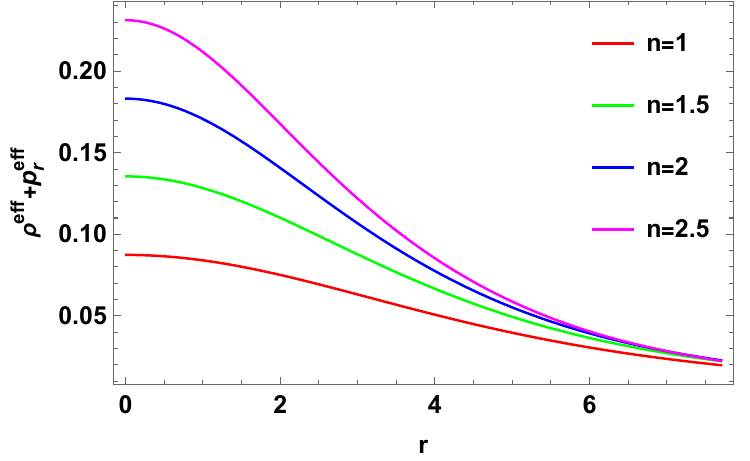}}
\subfloat[]{\includegraphics[width=70mm]{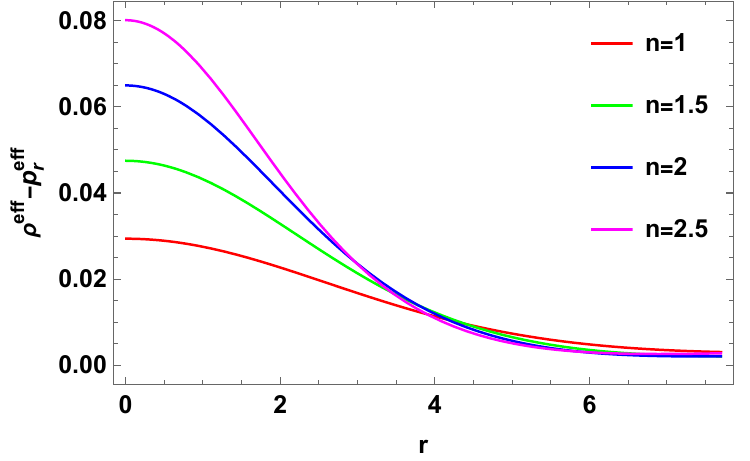}}
\\
\subfloat[]{\includegraphics[width=70mm]{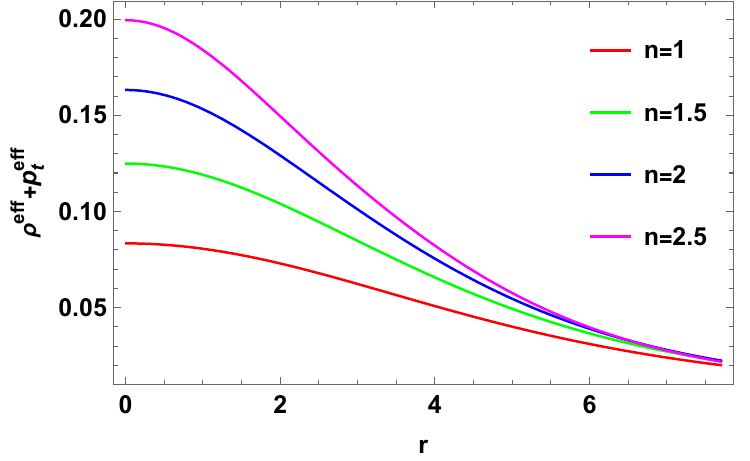}}
\subfloat[]{\includegraphics[width=70mm]{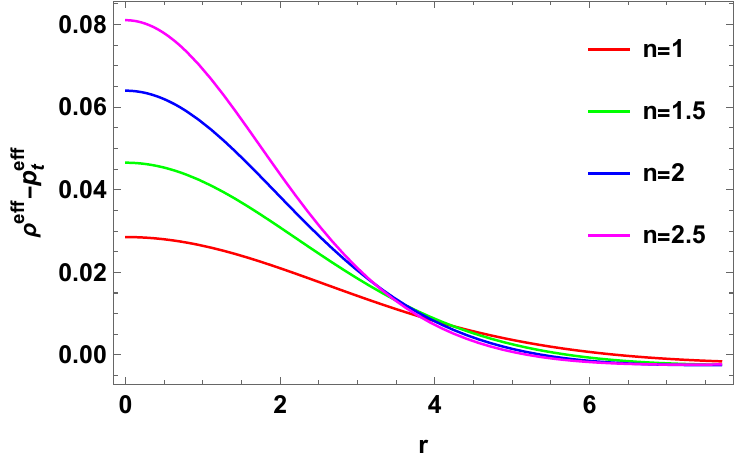}}
\\
\subfloat[]{\includegraphics[width=70mm]{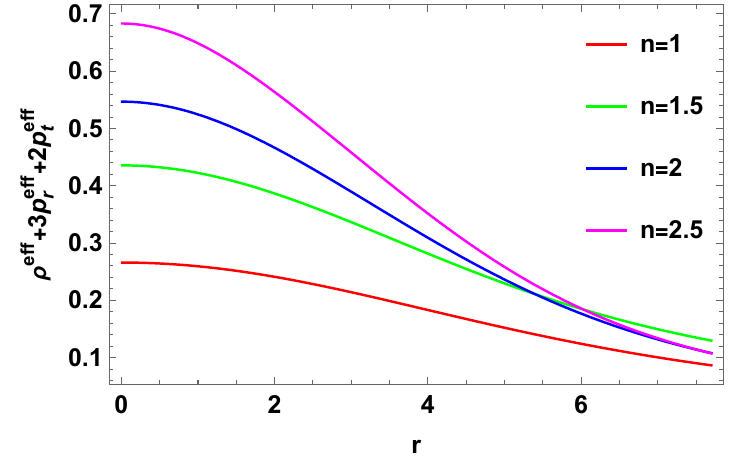}}
\hspace{200mm}
\caption{Graph of energy conditions $(km^{-2})$ with respect to radial variation $r~(km)$ for $ Her~ X-1$ through various GTK parametric values $n=1,~1.5,~2,~2.5$.}
\label{fig:8a-8c}
\end{figure}

The Fig (\ref{fig:8a-8c}) demonstrates that energy conditions are satisfied within the GTK metric on $f(R)$ framework, affirming the physical plausibility of the star $Her~ X -1$.

\subsection{The hydrostatic equilibrium condition}
The Tolman-Oppenheimer-Volkoff (TOV) equation is used to examine the star's stability under various forces. This version of the TOV equation, formulated by Ponce de León~\cite{69,Ponce}, accounts for the balancmodel's stability$(F_g)$, hydrostatic $(F_h)$, and anisotropic $(F_a)$ forces inside the star. It ensures that the outward pressure gradients and the effects of pressure anisotropy exactly counter the inward gravitational pull. Examining this balance is crucial to confirm that the stellar configuration remains in stable equilibrium throughout its interior \cite{Oppenheimer}.
\begin{equation}\label{TOV}
    -\frac{M_G(r)e^{\lambda-\nu}}{r^2}(\rho^{eff}+p_r^{eff})-\frac{dp_r^{eff}}{dr}+\frac{2\Delta}{r}=0.
\end{equation}
Here, $M_G$ represents the active gravitational mass given in \cite{1985Whittaker} as
\begin{equation}\label{Whittaker}
   M_G(r)=r^2\nu^{\prime} e^{\nu-\lambda}.
\end{equation}
Eq. (\ref{Whittaker}) is substituted in Eq. (\ref{TOV}), yields
\begin{equation}\label{TOV*}
   \frac{ -\nu^{\prime}}{2}(\rho^{eff}+p_r^{eff})-\frac{dp_r^{eff}}{dr}+\frac{2\Delta}{r}=0.
\end{equation}
Eq.~(\ref{TOV*}) incorporates the effects of forces $F_h$, $F_a$, and $F_g$
to represent the stability of the model.
    \begin{eqnarray}
    F_g&=&  \frac{ -\nu^{\prime}}{2}(\rho^{eff}+p_r^{eff}),\label{fg}\\
    F_h&=&-\frac{dp_r^{eff}}{dr},\label{fh}\\
    F_a&=&\frac{2\Delta}{r},\label{fa}
\end{eqnarray}
and
 \begin{eqnarray}
    F_g+F_h+F_a=0.
 \end{eqnarray}
The differences in the forces \( F_g \), \( F_h \), and \( F_a \) for \( n = 1 \), \( 1.5 \), \( 2 \), and \( 2.5 \) are depicted in Figs. (9a-9d), respectively. The graphs demonstrate that $F_g$ counteracts the combined influence of $F_h$ and $F_a$, maintaining the stability of $ Her~X-1$.

\begin{figure}[!htbp]  
\centering
\subfloat[]{\includegraphics[width=70mm]{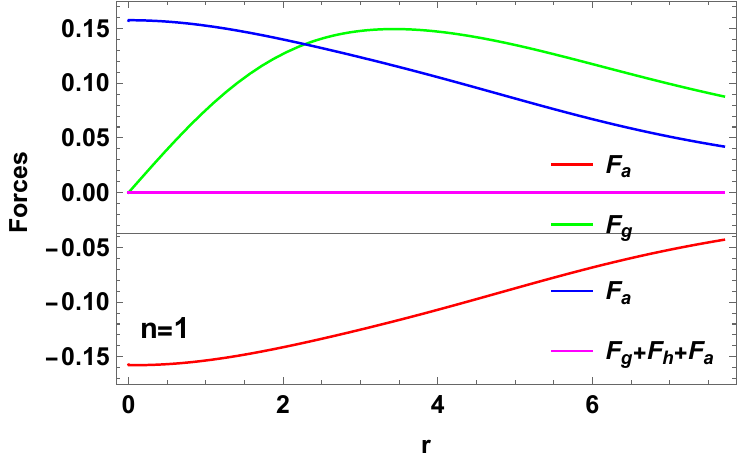}}
\subfloat[]{\includegraphics[width=70mm]{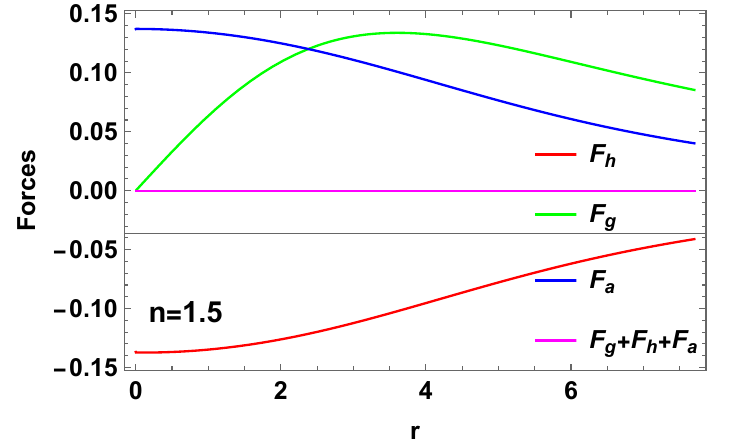}}
\\
\subfloat[]{\includegraphics[width=70mm]{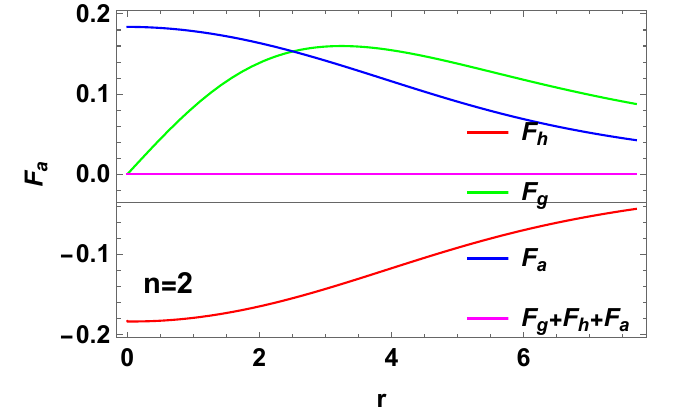}}
\subfloat[]{\includegraphics[width=70mm]{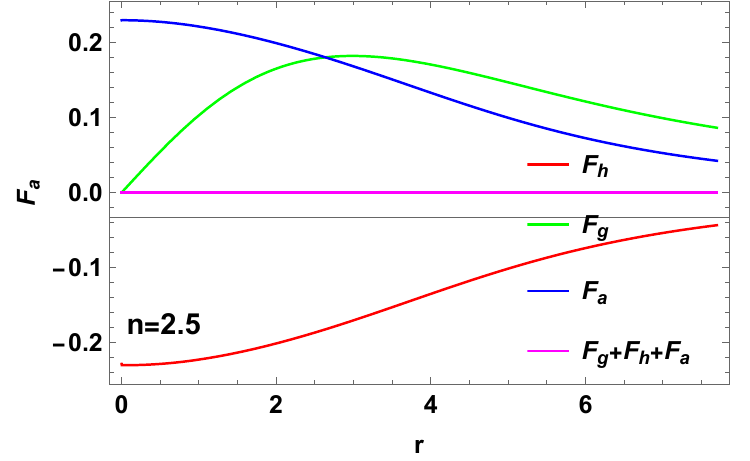}}
\hspace{200mm}
\caption{Graph of forces for $ Her~ X-1$ with respect to radial variation $r~(km)$ through various GTK parametric values $n=1,~1.5,~2,~2.5$. }
\label{fig:9}
\end{figure}
\subsection{Adiabatic index}
\begin{figure}
    \centering
    \includegraphics[width=0.60\linewidth]{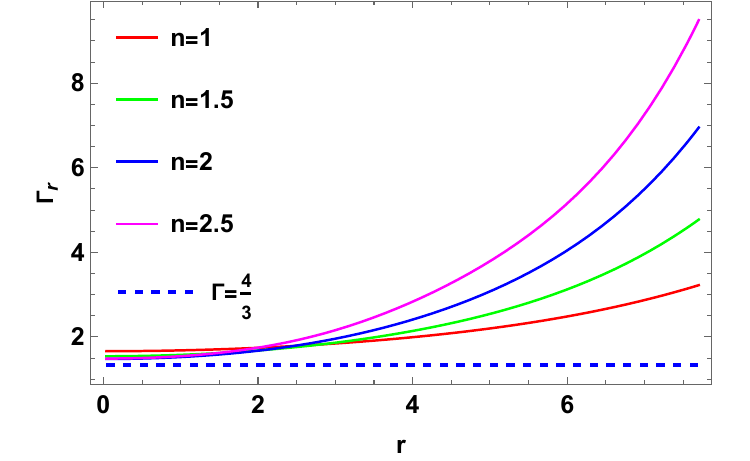}
    \caption{Graph of $\Gamma_r$ with respect to radial variation $r~(km)$ for $ Her~ X-1$ through various GTK parametric values $n=1,~1.5,~2,~2.5$.}
    \label{fig:10}
\end{figure}

The adiabatic index (\( \Gamma \)), which represents the specific heat ratio, is a fundamental parameter to understand the EoS at varying densities. Bondi~\cite{bondi1964} identified \( \Gamma_r \) as an essential tool for evaluating the response of anisotropic fluid spheres to small radial adiabatic perturbations.
\\
According to Newtonian theory, a stellar configuration remains stable if \( \Gamma > \frac{4}{3} \), while \( \Gamma = \frac{4}{3} \) corresponds to the condition for neutral equilibrium, as noted in Bondi’s work~\cite{bondi1964}. However, in relativistic contexts, this stability condition is modified because of additional pressure effects that can lead to significant instabilities. In general relativistic anisotropic systems, the presence of anisotropy further impacts stability, often requiring \( \Gamma \) exceeding \( \frac{4}{3} \) for the system to remain dynamically stable. This requirement is supported by the findings of Chan~\cite{1993}, Heinzmann~\cite{1967}, and Hillebrandt~\cite{1976}.
\\
The model considered here provides specific expressions for the radial and tangential adiabatic indices, given as 
\begin{equation}
    \Gamma_r=\Big(\frac{\rho+p_r}{p_r}\Big)v_r^2.
\end{equation}
As shown in Fig. (\ref{fig:10}), adiabatic indices higher than $\frac{4}{3}$ provide a stable configuration against the radial distribution.

\FloatBarrier
\subsection{Causality conditions and Herrera cracking technique}
\begin{figure}
    \centering
    \includegraphics[width=0.60\linewidth]{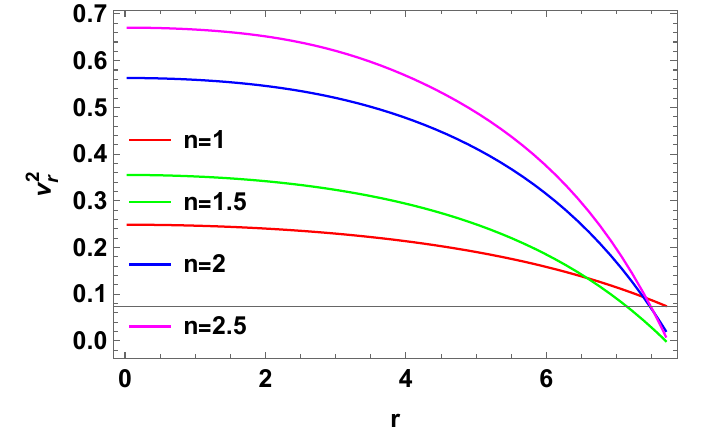}
    \caption{Graph of $v_r^2$ with respect to radial variation $r~(km)$ for $ Her~ X-1$ through various GTK parametric values $n=1,~1.5,~2,~2.5$.}
    \label{fig:11}
\end{figure}

\begin{figure}
    \centering
    \includegraphics[width=0.60\linewidth]{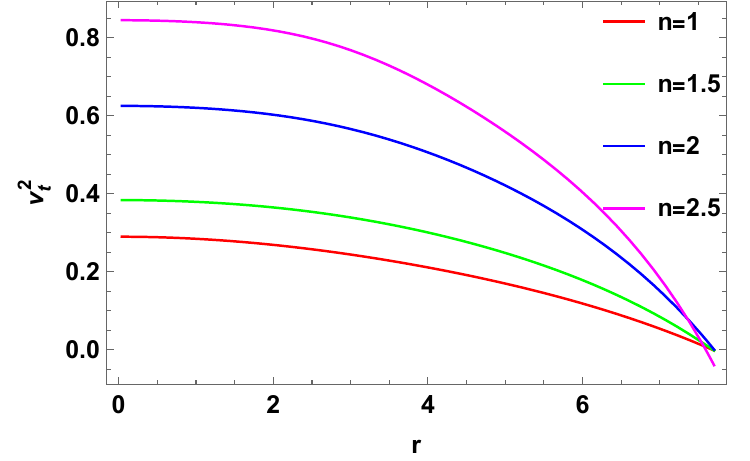}
    \caption{Graph of $v_t^2$ with respect to radial variation $r~(km)$ for $ Her~ X-1$ through various GTK parametric values $n=1,~1.5,~2,~2.5$.}
    \label{fig:12}
\end{figure}

\begin{figure}
    \centering
    \includegraphics[width=0.60\linewidth]{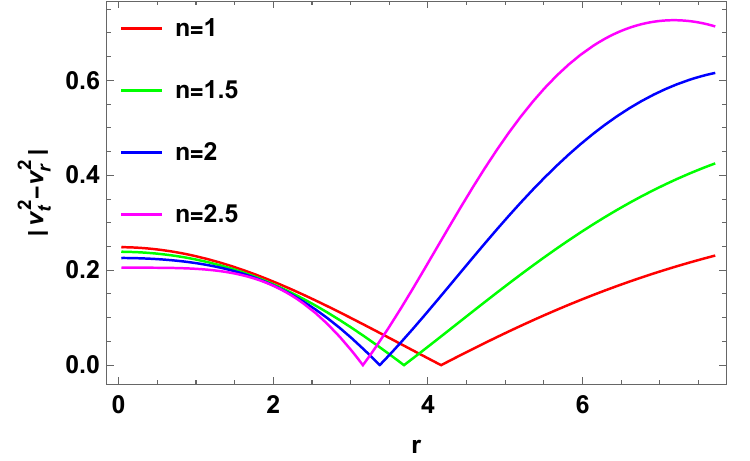}
    \caption{Graph of $|v_t^2-v_r^2|$ with respect to radial variation $r~(km)$ for $ Her~ X-1$ through various GTK  parametric values $n=1,~1.5,~2,~2.5$.}
    \label{fig:13}
\end{figure}

For a self-gravitating system to be physically viable, the causality conditions must be satisfied. The radial (\(v_r^2\)) and the tangential (\(v_t^2\)) sound speed must be within the interval \( [0, 1] \). This criterion guarantees the consistency of the causality by preventing sound from traveling faster than light~\cite{64}-\cite{59}.
\\
In addition, the stability analysis incorporates Abreu’s criterion in conjunction with Herrera’s cracking concept, mathematically stated as
\[
0 < \lvert v_t^2 - v_r^2 \rvert < 1.
\]
To assess the model's stability, the profiles of \( v_r^2 \) and \( v_t^2 \) are examined, as illustrated in Figs.~(\ref{fig:11}) and~(\ref{fig:12}), respectively. The analysis demonstrates that both \( v_r^2 \) and \( v_t^2 \) remain within the range \( [0, 1] \) throughout the star interior, thus upholding the causality condition ensuring the physical acceptability of the model. Fig.~(\ref{fig:13}) shows that \(| v_t^2 - v_r^2| \) remains positive throughout the stellar interior for all values of \( n \), indicating that the transverse sound speed exceeds the radial one. This behavior satisfies Abreu’s stability criterion, suggesting a stable configuration for $Her~ X-1$.
\begin{figure}
    \centering
    \includegraphics[width=0.5\linewidth]{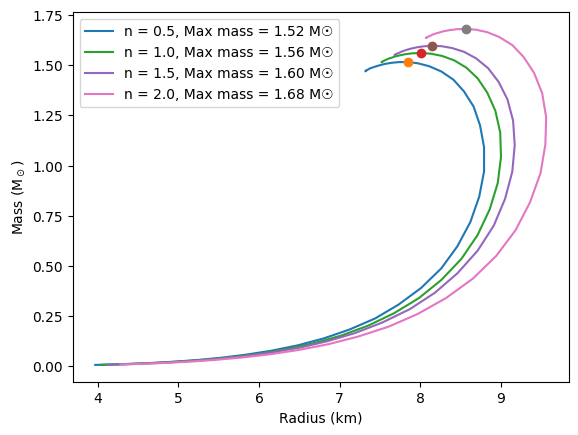}
    \caption{Graph of mass-radius relation of $Her~X-1$ for different values of $n$}
    \label{fig:14}
\end{figure}

\section{Conclusion and Discussion}

In this study, we have explored the dynamical evolution of a self-gravitational compact object influenced by the presence of exotic matter within the framework of modified gravity, adopting the Starobinsky model $f(R) = R + \alpha R^2$, which introduces higher-order curvature corrections to account for nonstandard gravitational effects. The model inherently leads to a system of nonlinear differential equations (DEs) due to the quadratic curvature term, which significantly complicates the structure equations governing the stellar system. We investigate the nonlinear spherical evolution of stellar configurations, particularly under the influence of anisotropic pressure distributions, a realistic feature in compact astrophysical objects.
To manage the resulting nonlinearity, we employ a combination of mathematical techniques. First, we have taken a static, spherically symmetric spacetime using a GTK metric, in which the problem becomes only radially dependent. This metric ansatz facilitates a tractable formulation of the modified FEs. Secondly, by assuming a specific form of the energy-momentum tensor that includes anisotropic stresses and DM contributions, we have reduced the system to a set of coupled nonlinear ordinary DEs. In cases where exact analytical solutions are infeasible, the equations have been solved numerically using feasible boundary conditions at the stellar center and the boundary surface.

Our results have demonstrated that as the GTK metric parameter $n$ increases from 1 to 2.5, the central density exhibits a pronounced enhancement in Fig. (\ref{fig:1}). At the same time, the overall distribution remains elevated across a broader radial extent. This behavior indicates a strong correlation between the parameter $n$ and the concentration of DM within the stellar configuration. Higher values of $n$ yield denser and more spatially extended DM distributions, thereby amplifying its dynamical contribution to the system. These results underscore the significant role of the parameter $n$ in modulating the DM profile and its influence on the internal structure of compact stars.

The studies on the effects of $p_r^{{eff}}$ in the case of the stellar object $Her~X-1$ presented in Fig. (\ref{fig:2}). We have found that as $n$ increases from 1 to 2.5, the central pressure (at $r = 0$) rises significantly, and the pressure remains elevated throughout the stellar interior, indicating the presence of DM, modeled through the parameter $n$, contributes to a stiffer state and provides additional pressure support against gravitational collapse. The enhancement in $p_r^{{eff}}$ with increasing $n$ underscores the growing dynamical influence of DM in maintaining hydrostatic equilibrium. Thus, a stronger internal counterforce stabilizes the stellar configuration.

From the graphical analyses of the effective pressure $p_t^{eff}$, we have shown that in Fig. (\ref{fig:3}) higher values of \( n \) correspond to greater values of pressure throughout the radial distance. The pressure peaks at the center and gradually decreases outward, but remains higher overall as \( n \) increases, leading to a more pronounced tangential pressure contribution that may reflect stronger anisotropic effects induced by the presence of DM. Such behavior highlights the role of DM in shaping the internal structure and supporting the system against instabilities through enhanced tangential stresses.

The radial variation of the anisotropy \( \Delta \) for Her~X--1 in Fig. (\ref{fig:4}) starts from zero in the center and increases outward at larger radii. As \( n \) increases, the magnitude of anisotropy becomes more pronounced across the entire stellar configuration. This trend suggests that higher values of \( n \) enhance the disparity between tangential and radial pressures, leading to a stronger anisotropic force within the system. This behavior shows that DM not only influences the density and pressure profiles but also plays a key role in generating and sustaining anisotropic effects, which can contribute to the overall structural balance and stability of the system.

Additionally, we have found that the derivatives of both effective pressure and density are negative throughout the radial coordinate, as shown in Figs. (\ref{fig:5}-\ref{fig:7}). It means that a monotonic decrease in energy density and pressure from the center toward the surface of the stellar configuration. At larger values of $n$, the magnitude of the gradients also increases. It implies that the quantities fall off more sharply with radius for larger $n$. 
On the other hand, higher values of $n$ make the object more compact and denser structures, as confirmed with the influence of DM on the pressure and energy density. For larger \( n \), also suggest enhanced internal forces needed to counterbalance gravity, maintaining equilibrium in the presence of dominant DM effects.

We have obtained results demonstrating that the NECs decrease with radius but remain positive for all values of $n$, thereby satisfying the NECs. Similarly, we have shown that the DECs exhibit a dip while staying positive across the entire radial profile, fulfilling the condition for all $n$. Furthermore, we have found that the SEC decreases with $r$ but remains positive, indicating that the SEC is met within the stellar system. These ECs are illustrated in Figs. (\ref{fig:8a-8c})
We have analyzed the different forces, including $F_g$, $F_h$, and $F_a$, for various values of $n$ in Fig. (\ref{fig:9}). As $n$ increases, we have observed that the magnitude and steepness of all forces intensify, reflecting more compact and dense stellar structures. We have determined that the hydrostatic force acts outward to counterbalance the inward gravitational pull, while the anisotropic force grows with radius, indicating increasing pressure anisotropy in outer regions. The enhanced gradients for higher $n$ values suggest a stronger role of internal forces in maintaining balance, which aligns with the influence of DM. Its presence effectively amplifies the overall gravitational pull, necessitating greater hydrostatic and anisotropic responses to preserve equilibrium. Thus, we have illustrated how the variation in force profiles with $n$ highlights the potential contribution of DM to the internal dynamics and structural stability of compact stellar systems.
We have examined the radial adiabatic index $\Gamma_r$ versus radius $r$ for different values of $n$ in Fig. (\ref{fig:10}). As $n$ increases from 1 to 2.5, we have found that $\Gamma_r$ rises more rapidly, indicating greater stiffness and stability in stellar matter. All profiles lie above the critical line $\Gamma_r = \frac{4}{3}$, confirming dynamic stability throughout the star. The increase in $\Gamma_r$ with higher $n$ suggests a stronger internal pressure response, possibly influenced by DM, that helps prevent gravitational collapse.
We have investigated the behavior of $v_r^2$ and $v_t^2$, along with their absolute difference $|v_t^2 - v_r^2|$, as functions of radius $r$ for varying values of the parameter $n$ in Figs. (\ref{fig:11}-\ref{fig:13}). Our results reveal that $v_r^2$ increases with $r$ and becomes steeper as $n$ rises, indicating a stronger radial response in more compact configurations. A similar trend has been observed for $v_t^2$, reflecting the tangential sound speed’s sensitivity to increasing anisotropy at higher $n$. The difference $|v_t^2 - v_r^2|$ remains small near the center and increases outward, particularly for larger $n$, confirming that pressure anisotropy becomes more significant in the outer region. These findings suggest enhanced anisotropic behavior and sound speed dispersion for higher $n$, possibly due to DM effects or structural stiffness in more compact stars.  Fig. (\ref{fig:14}) illustrates how the mass of $Her~X–1$ varies with its radius for different values of the parameter $n$. This relation provides insights into the structural properties of the star under varying model conditions.
  Each curve corresponds to the dot that marks the maximum stable mass, representing the observational constraints $M \approx 1.5 \pm 0.1\,M_{\odot}$, showing good agreement with the theoretical predictions.
 
A comparative analysis is performed between our results and those reported in the literature under $f(R)$ gravity. In particular, we have referred to the works of Astashenok et al. \cite{Astashenok}, Zubair and Abbas \cite{zubair2014}, who explored compact stars in \( f(R) = R + \alpha R^2 \) gravity and reported that higher-order curvature corrections significantly affect stellar properties, including mass-radius relations, energy density profiles, and stability indicators. Our results are consistent with existing studies. Specifically, the increase in the parameter \( n \) in the GTK metric leads to enhanced central density, sharper pressure gradients, and stronger anisotropy, effects that reflect increased contributions of curvature analogous to the impact of higher values of \( \alpha \) in the Starobinsky model. Moreover, the fulfillment of the energy conditions and the stability ensured by the adiabatic index \( \Gamma_r > \frac{4}{3} \) are consistent with the criteria established in the literature mentioned above.
These comparisons have been incorporated into the discussion and conclusion sections to contextualize our results within \( f(R) \) gravity.
Our results show that the nonlinear nature of the theory significantly affects the distribution and influence of DM within the star. Higher-order curvature terms dynamically couple with the DM distribution, modifying the radial and tangential pressure gradients. This coupling improves internal pressure support against gravitational collapse and maintains hydrostatic equilibrium. The model is applied to a test compact object from the Hercules constellation, $ Her~X-1$, to determine the physical viability of the solution. We show that for increasing values of the parameter $n$ associated with the GTK metric, the physical quantities show singularity-free behavior. Moreover, the parameter $n$ critically influences the stiffness and internal dynamics of the configuration, highlighting its role in determining the structural stability of this system in $f(R)$ gravity.

Future work will involve constructing new models in other modified gravity theories with similar configurations, which will contrast energy–condition diagnostics and stability with our stellar models. A natural extension is to examine whether the higher–order curvature sector in Starobinsky $f(R)$ can self–support a traversable wormhole throat by adopting a wormhole ansatz and scanning the space for regimes where curvature provides the effective exoticity. Recent $f(T)$ studies indicate that modified geometric terms can sustain traversable wormholes without ordinary exotic matter, suggesting an analogous possibility in $f(R)$ under suitable boundary data and matching conditions \cite{Ainamon2024,Tefo2019}. Similarly, further extensions will be development of the model in different gravitational theories like $f(R,T)$, $f(R, L_m)$ and $f(Q,T)$  \cite{Yerlan, Koussour, Zhadyranova}.
\section*{Data availability statement}
This is a theoretical study, so no data is associated with it. 


\begin{thebibliography}{99}

\bibitem{1} Ade, P. A., et al. (2014). Planck 2013 results. I. Overview of products and scientific results. Astronomy \& Astrophysics, 571, A1.
\bibitem{kamenshchik2001alternative} Kamenshchik, A., Moschella, U., and Pasquier, V. (2001). An alternative to quintessence. Physics Letters B, 511(2-4), 265-268.
\bibitem{padmanabhan2002can} Padmanabhan, T., and Choudhury, T. R. (2002). Can the clustered dark matter and the smooth dark energy arise from the same scalar field?. Physical Review D, 66(8), 081301.
\bibitem{bento2002generalized} Bento, M. C., Bertolami, O., and Sen, A. A. (2002). Generalized Chaplygin gas, accelerated expansion, and dark-energy-matter unification. Physical Review D, 66(4), 043507.
\bibitem{caldwell2002phantom} Caldwell, R. R. (2002). A phantom menace? Cosmological consequences of a dark energy component with super-negative equation of state. Physics Letters B, 545(1-2), 23-29.
\bibitem{nojiri2003quantum} Nojiri, S. I., and Odintsov, S. D. (2003). Quantum de Sitter cosmology and phantom matter. Physics Letters B, 562(3-4), 147-152.
\bibitem{nojiri2003sitter} Nojiri, S. I., and Odintsov, S. D. (2003). de Sitter brane universe induced by phantom and quantum effects. Physics Letters B, 565, 1-9.
\bibitem{riess2004type} Riess, A. G., Strolger, L. G., Tonry, J., Casertano, S., Ferguson, H. C., Mobasher, B., and Tsvetanov, Z. (2004). Type Ia supernova discoveries at $z> 1$ from the Hubble Space Telescope: Evidence for past deceleration and constraints on dark energy evolution. The Astrophysical Journal, 607(2), 665.
\bibitem{rakhmanov2025optical} Rakhmanov, S., Matchonov, K., Yusupov, H., Nasriddinov, K., and Matrasulov, D. (2025). Optical high harmonic generation in Dirac materials. The European Physical Journal B, 98(2), 35.
\bibitem{kobayashi2008relativistic} Kobayashi, T.,  and Maeda, K. I. (2008). Relativistic stars in $f (R)$ gravity, and absence thereof. Physical Review D, 78(6), 064019.
\bibitem{kobayashi2009can} Kobayashi, T., and Maeda, K. I. (2009). Can higher curvature corrections cure the singularity problem in $f (R)$ gravity?. Physical Review D, 79(2), 024009.
\bibitem{babichev2009relativistic} Babichev, E., and Langlois, D. (2009). Relativistic stars in $f (R)$ gravity. Physical Review D, 80(12), 121501.
\bibitem{upadhye2009existence} Upadhye, A., and Hu, W. (2009). Existence of relativistic stars in $f (R)$ gravity. Physical Review D, 80(6), 064002.
\bibitem{nzioki2010new} Nzioki, A. M., Carloni, S., Goswami, R., and Dunsby, P. K. (2010). New framework for studying spherically symmetric static solutions in $f (R)$ gravity. Physical Review D, 81(8), 084028.
\bibitem{kausar2014dissipative} Kausar, H. R., and Noureen, I. (2014). Dissipative spherical collapse of charged anisotropic fluid in $f(R)$  gravity. The European Physical Journal C, 74, 1-8.
\bibitem{rizwana2015dynamical} Rizwana Kausar, H., Noureen, I., and Shahzad, M. U. (2015). Dynamical analysis of charged anisotropic spherical star in $f(R)$ gravity. The European Physical Journal Plus, 130(10), 204.
\bibitem{naz2021embedded} Naz, T., Usman, A., and Shamir, M. F. (2021). Embedded class-I solution of compact stars in $f(R)$ gravity with Karmarkar condition. Annals of Physics, 429, 168491.
\bibitem{nashed2021anisotropic} Nashed, G. G. L., and Capozziello, S. (2021). Anisotropic compact stars in $f(R)$ gravity. The European Physical Journal C, 81(5), 1-20.
\bibitem{nojiri2007introduction} Nojiri, S. I., and Odintsov, S. D. (2007). Introduction to modified gravity and gravitational alternative for dark energy. International Journal of Geometric Methods in Modern Physics, 4(01), 115-145.
\bibitem{Manzoor} Manzoor, R., and Shahid, W. (2021). Evolution of cluster of stars in $f(R)$ gravity. Physics of the Dark Universe, 33, 100844.
\bibitem{Praagman} Praagman, A., Hurley, J., and Power, C. (2010). Star cluster evolution in dark matter dominated galaxies. New Astronomy, 15(1), 46–51. 
\bibitem{Baumgardt} Baumgardt, H., and Makino, J. (2003). Dynamical evolution of star clusters in tidal fields. Monthly Notices of the Royal Astronomical Society, 340(1), 227-246.
\bibitem{Shahid} Shahid, W., Manzoor, R., Mumtaz, S., Mardan, S. A., and Malik, A. (2024). Stability of evolving cluster of stars and exotic matter. The European Physical Journal C, 84(12), 1336.
\bibitem{Manzoor1} Manzoor, R., Jawad, A., Adeel, M., Saeed, M., and Rani, S. (2019). Collapsing stellar filament and exotic matter in Palatini $f(R)$ gravity. The European Physical Journal C, 79(10), 831.
\bibitem{Starobinsky} Starobinsky, A. A. (1980). A new type of isotropic cosmological models without singularity. Physics Letters B, 91(1), 99–102.
\bibitem{Percacci} Percacci, R., and Vacca, G. P. (2025). The Starobinsky model of inflation and Renormalizability. arXiv preprint arXiv:2502.13931.
\bibitem{Naseer} Naseer, T., and Sharif, M. (2024). Charged anisotropic Starobinsky models admitting vanishing complexity. Physics of the Dark Universe, 46, 101595.
\bibitem{Gialamas} Gialamas, I. D., and Tamvakis, K. (2023). Bimetric starobinsky model. Physical Review D, 108(10), 104023.
\bibitem{Ivanov} Ivanov, V. R., Ketov, S. V., Pozdeeva, E. O., and Vernov, S. Y. (2022). Analytic extensions of Starobinsky model of inflation. Journal of Cosmology and Astroparticle Physics, 2022(03), 058.
\bibitem{Cheong} Cheong, D. Y., Lee, H. M., and Park, S. C. (2020). Beyond the Starobinsky model for inflation. Physics Letters B, 805, 135453.
\bibitem{Aldabergenov} Aldabergenov, Y., Ishikawa, R., Ketov, S. V., and Kruglov, S. I. (2018). Beyond Starobinsky inflation. Physical Review D, 98(8), 083511.
\bibitem{Sekhmani} Sekhmani, Y., Gashti, S. N., Afshar, M. A. S., Alipour, M. R., Sadeghi, J., Pourhassan, B., and Rayimbaev, J. (2024). Thermodynamic topology of Black Holes in $ F (R)$ Euler-Heisenberg gravity's Rainbow. arXiv preprint arXiv:2409.04997.
\bibitem{Oppenheimer} Oppenheimer, J. R., and Volkoff, G. M. (1939). On massive neutron cores. Physical Review, 55(4), 374.
\bibitem{martin2004algorithmic} Martin, D., and Visser, M. (2004). Algorithmic construction of static perfect fluid spheres. Physical Review D, 69(10), 104028.
\bibitem{boonserm2005generating} Boonserm, P., Visser, M., and Weinfurtner, S. (2005). Generating perfect fluid spheres in general relativity. Physical Review D, 71(12), 124037.
\bibitem{sarkar2019compact} Sarkar, N., Singh, K. N., Sarkar, S., and Rahaman, F. (2019). Compact star models in class I spacetime. The European Physical Journal C, 79(6), 516.
\bibitem{singh2019generalized} Singh, K. N., Maurya, S. K., Rahaman, F., and Tello-Ortiz, F. (2019). A generalized Finch–Skea class one static solution. The European Physical Journal C, 79(5), 381.
\bibitem{tello2019anisotropic} Tello-Ortiz, F., Maurya, S. K., Errehymy, A., Singh, K. N., and Daoud, M. (2019). Anisotropic relativistic fluid spheres: an embedding class I approach. The European Physical Journal C, 79(11), 885.
\bibitem{maurya2020anisotropic} Maurya, S. K., Newton Singh, K., Errehymy, A., and Daoud, M. (2020). Anisotropic stars in $f ({\textit {G}},{\textit {T}})$ gravity under class I space-time. The European Physical Journal Plus, 135(10), 1-20.
\bibitem{maurya2020non} Maurya, S. K., Singh, K. N., and Dayanandan, B. (2020). Non-singular solution for anisotropic model by gravitational decoupling in the framework of complete geometric deformation (CGD). The European Physical Journal C, 80(5), 448.
\bibitem{singh2020compact} Singh, K. N., Ali, A., Rahaman, F., and Nasri, S. (2020). Compact stars with exotic matter. Physics of the Dark Universe, 29, 100575.
\bibitem{das2021modeling} Das, S., Singh, K. N., Baskey, L., Rahaman, F., and Aria, A. K. (2021). Modeling of compact stars: an anisotropic approach. General Relativity and Gravitation, 53(3), 25.
\bibitem{zubair2021realistic} Zubair, M., Waheed, S., Jamal, M. F., and Mustafa, G. (2021). Realistic anisotropic Karmarkar stars in Rastall gravitational framework. Results in Physics, 29, 104787.
\bibitem{tamta2022study}  Tamta, P., and Fuloria, P. (2022). Study of anisotropic stellar objects, a revisit to Buchdahl metric potential. International Journal of Modern Physics D, 31(08), 2250057.
\bibitem{kumar2021generalized}  Kumar, J., Singh, H. D., and Prasad, A. K. (2021). A generalized Buchdahl model for compact stars in $f (R, T)$ gravity. Physics of the Dark Universe, 34, 100880.
\bibitem{maurya2019anisotropic}  Maurya, S. K., Banerjee, A., Jasim, M. K., Kumar, J., Prasad, A. K., and Pradhan, A. (2019). Anisotropic compact stars in the Buchdahl model: A comprehensive study. Physical Review D, 99(4), 044029.
\bibitem{maurya2018role} Maurya, S. K., Banerjee, A., and Hansraj, S. (2018). Role of pressure anisotropy on relativistic compact stars. Physical Review D, 97(4), 044022.
\bibitem{singh2020anisotropic}  Singh, K. N., Maurya, S. K., Bhar, P., and Rahaman, F. (2020). Anisotropic stars with a modified polytropic equation of state. Physica Scripta, 95(11), 115301.
\bibitem{akramov2024discrete} Akramov, M., Trunk, C., Yusupov, J., and Matrasulov, D. (2024). Discrete Schrödinger equation on graphs: an effective model for branched quantum lattice. Europhysics Letters, 147(6), 62001.
\bibitem{maurya2019effect} Maurya, S. K., Maharaj, S. D., Kumar, J., and Prasad, A. K. (2019). Effect of pressure anisotropy on Buchdahl-type relativistic compact stars. General Relativity and Gravitation, 51(7), 86.
\bibitem{gabbanelli2018gravitational} Gabbanelli, L., Rincón, Á., and Rubio, C. (2018). Gravitational decoupled anisotropies in compact stars. The European Physical Journal C, 78(5), 370.
\bibitem{turimov2025particles} Turimov, B., Usanov, S., and Khamroev, Y. (2025). Particles acceleration by Bocharova–Bronnikov–Melnikov–Bekenstein black hole. Physics of the Dark Universe, 48, 101876.
\bibitem{gabbanelli2019causal} Gabbanelli, L., Ovalle, J., Sotomayor, A., Stuchlik, Z., and Casadio, R. (2019). A causal Schwarzschild-de Sitter interior solution by gravitational decoupling. The European Physical Journal C, 79(6), 486.
\bibitem{lopes2019anisotropic} Lopes, I., Panotopoulos, G., and Rincón, Á. (2019). Anisotropic strange quark stars with a nonlinear equation-of-state. The European Physical Journal Plus, 134(9), 454.
   \bibitem{26} Biswas, S., Shee, D., Ray, S., Rahaman, F., and Guha, B. K. (2019). Relativistic strange stars in Tolman–Kuchowicz spacetime. Annals of Physics, 409, 167905.
    \bibitem{27} Jasim, M. K., Deb, D., Ray, S., Gupta, Y. K., and Chowdhury, S. R. (2018). Anisotropic strange stars in Tolman–Kuchowicz spacetime. The European Physical Journal C, 78(7), 603.
    \bibitem{28} Bhar, P., Singh, K. N., and Tello-Ortiz, F. (2019). Compact star in Tolman–Kuchowicz spacetime in the background of Einstein–Gauss–Bonnet gravity. The European Physical Journal C, 79(11), 922.
    \bibitem{29} Acharya, H. R., Pandya, D. M., Parekh, B., and Thomas, V. O. (2025). Relativistic compact object in Generalised Tolman-Kuchowicz spacetime with quadratic equation of state. arXiv preprint arXiv:2504.02311.
 
    \bibitem{nazar2025possible} Nazar, H., Abbas, G., Shahzad, M. R., Ashraf, A., Boukhris, I., Alanazi, A. A., and Atamurotov, F. (2025). Possible existence of traversable wormholes within stellar galactic halos in modified $ f (R)$ gravity: A class 1 embedding approach. Physics of the Dark Universe, 48, 101837.
    \bibitem{nazar2025exhibiting} Nazar, H., Majeed, A., Abbas, G., Ashraf, A., and Channuie, P. (2025). Exhibiting stable model of dark energy compact star with Tolman-VI solution under complexity free system. The European Physical Journal C, 85(2), 125.
    \bibitem{aslam2025impact} Aslam, M., and Malik, A. (2025). Impact of Tolman–Kuchowicz solution on dark energy compact stars in $f (R)$ theory. Annals of Physics, 472, 169854.
    \bibitem{malik2025modeling} Malik, A., Aslam, M., Chaudhary, S., Almas, A., and Kausar, G. (2025). Modeling of Hybrid Baryonic-Quark Matter in $f (R)$ Gravity with scalar potential. Annals of Physics, 473, 169896.
   \bibitem{56} Das, B., Das, S., and Paul, B. C. (2023). Models of compact objects with charge in generalized Tolman-Kuchowicz metric. Astrophysics and Space Science, 368(11), 98.
    \bibitem{44} Tolman, R. C. (1939). Static solutions of Einstein's field equations for spheres of fluid. Physical Review, 55(4), 364.
    \bibitem{45} Kuchowicz, B. (1968). General relativistic fluid spheres. I. New solutions for spherically symmetric matter distributions. Acta Physica Polonica, 33, 541–563.
    \bibitem{ruderman1972pulsars} Ruderman, M. (1972). Pulsars: structure and dynamics. Annual Review of Astronomy and Astrophysics, 10, 427.
    \bibitem{glendenning2012compact} Glendenning, N. K. (2012). Compact stars: Nuclear physics, particle physics and general relativity. Springer Science \& Business Media.
    \bibitem{herzog2011three} Herzog, M., and Röpke, F. K. (2011). Three-dimensional hydrodynamic simulations  format of the combustion of a neutron star into a quark star. Physical Review D, 84(8), 083002.
    \bibitem{zel2014stars} Zel’dovich, Y. B., and Novikov, I. D. (2014). Stars and relativity. Courier Corporation.
    \bibitem{misner1964relativistic} Misner, C. W., and Sharp, D. H. (1964). Relativistic equations for adiabatic, spherically symmetric gravitational collapse. Physical Review, 136(2B), B571.
    \bibitem{shee2017compact} Shee, D., Ghosh, S., Rahaman, F., Guha, B. K., and Ray, S. (2017). Compact star in pseudo-spheroidal spacetime. Astrophysics and Space Science, 362(6), 114.
    \bibitem{quevedo1989general} Quevedo, H. (1989). General static axisymmetric solution of Einstein’s vacuum field equations in prolate spheroidal coordinates. Physical Review D, 39(10), 2904.
    \bibitem{chifu2012gravitational} Chifu, E. N. (2012). Gravitational fields exterior to a homogeneous spherical masses. The Abraham Zelmanov Journal, 5, 31-67.
    \bibitem{deb} Deb, D., Chowdhury, S. R., Ray, S., Rahaman, F., and Guha, B. K. (2017). Relativistic model for anisotropic strange stars. Annals of Physics, 387, 239-252.
    \bibitem{astashenok1} Astashenok, A. V. (2014). Neutron star models in frames of $f (R)$ gravity. In AIP Conference Proceedings 1606(1), 99-104. American Institute of Physics.
    \bibitem{hossein} Hossein, S. M., Rahaman, F., Naskar, J., Kalam, M., and Ray, S. (2012). Anisotropic compact stars with variable cosmological constant. International Journal of Modern Physics D, 21(13), 1250088.
    \bibitem{zahra2025investigating} Zahra, A., Mardan, S. A., Riaz, M. B., and Kozubek, T. (2025). Investigating generalized polytropic compact objects in $f (R)$ gravity. The European Physical Journal C, 85(3), 291.
    \bibitem{schwarzschild1999gravitational} Schwarzschild, K. (1916). On the gravitational field of a mass point according to Einstein’s theory. Sitzungsberichte der Königlich Preussischen Akademie der Wissenschaften, 189.
    \bibitem{darmois1927equations} Darmois, G. (1927). Les équations de la gravitation einsteinienne, 58.
    \bibitem{israel1966singular} Israel, W. (1966). Singular hypersurfaces and thin shells in general relativity. Il Nuovo Cimento B, 44(1), 1-14.
    \bibitem{chu2022generalized} Chu, C. S., and Tan, H. S. (2022). Generalized darmois–israel junction conditions. Universe, 8(5), 250.
    \bibitem{abubekerov} Abubekerov, M. K., Antokhina, E. A., Cherepashchuk, A. M., and Shimanskii, V. V. (2008). The mass of the compact object in the X-ray binary her X-1/HZ her. Astronomy Reports, 52, 379-389.
    \bibitem{85} Bhar, P. (2019). Anisotropic compact star model: a brief study via embedding. The European Physical Journal C, 79, 1-13.
    \bibitem{87} Lattimer, J. M., and Steiner, A. W. (2014). Neutron star masses and radii from quiescent low-mass X-ray binaries. The Astrophysical Journal, 784(2), 123.
     \bibitem{leahy2025geometry} Leahy, D., and Mendelsohn, J. (2025). The geometry of the Hercules X-1 accretion disk from X-rays. Discover Space, 129(1), 5.
    \bibitem{kolassis1988energy} Kolassis, C. A., Santos, N. O., and Tsoubelis, D. (1988). Energy conditions for an imperfect fluid. Classical and Quantum Gravity, 5(10), 1329.
    \bibitem{hawking2023large} Hawking, S. W., and Ellis, G. F. (2023). The large scale structure of space-time. Cambridge university press.
    \bibitem{wald2010general} Wald, R. M. (2010). General relativity. University of Chicago press.
    \bibitem{brassel2021inhomogeneous} Brassel, B. P., Maharaj, S. D., and Goswami, R. (2021). Inhomogeneous and radiating composite fluids. Entropy, 23(11), 1400.
    \bibitem{brassel2021higher} Brassel, B. P., Maharaj, S. D., and Goswami, R. (2021). Higher-dimensional inhomogeneous composite fluids: energy conditions. Progress of Theoretical and Experimental Physics, 2021(10), 103E01.
    \bibitem{69} Ponce de Leon, J. (1987). General relativistic electromagnetic mass models of neutral spherically symmetric systems. General relativity and gravitation, 19, 797-807.
    \bibitem{Ponce} Ponce de Leon, J. (1993). Limiting configurations allowed by the energy conditions. General relativity and gravitation, 25(11), 1123-1137.
    
    
\bibitem{1985Whittaker}
Grøn, Ø. (1985). Repulsive gravitation and electron models. Physical Review D, 31(8), 2129.
 \bibitem{bondi1964} Bondi, H. (1964). The contraction of gravitating spheres. Proceedings of the Royal Society of London. Series A. Mathematical and Physical Sciences, 281(1384), 39-48.
    \bibitem{1993} Chan, R., Herrera, L., and Santos, N. O. (1993). Dynamical instability for radiating anisotropic collapse. Monthly Notices of the Royal Astronomical Society, 265(3), 533-544.
    \bibitem{1967} Heintzmann, H., and Hillebrandt, W. (1975). Neutron stars with an anisotropic equation of state: Mass, redshift, and stability. Astronomy and Astrophysics, 38(1), 51–55.
    \bibitem{1976} Hillebrandt, W., and Steinmetz, K. O. (1976). Anisotropic neutron star models: Stability against radial and nonradial pulsations. Astronomy and Astrophysics, 53(2), 283–287.
    \bibitem{64} Herrera, L. (1992). Cracking of self-gravitating compact objects. Physics Letters A, 165(3), 206-210.
    \bibitem{59} Abreu, H., Hernández, H., and Núnez, L. A. (2007). Sound speeds, cracking and the stability of self-gravitating anisotropic compact objects. Classical and Quantum Gravity, 24(18), 4631.
\bibitem{Astashenok} Astashenok, A. V., Capozziello, S., and Odintsov, S. D. (2013). Further stable neutron star models from $f (R)$ gravity. Journal of Cosmology and Astroparticle Physics, 2013(12), 040.  \bibitem{zubair2014} Zubair, M.,  Abbas, G. (2014). Study of Anisotropic Compact Stars in Starobinsky Model. arXiv:1412.2120v3.

\bibitem{Ainamon2024}
Aïnamon, C., Ganiou, M. G., Tefo, R. C., and  Houndjo, M. J. S. (2024). 
$f(T)$ theory solutions for traversable wormhole existence and neutron stars mass limits problems. 
International Journal of Geometric Methods and Modern Physics, 21, 2450036. 

\bibitem{Tefo2019}
 Tefo, R. C.,  Logbo, P. H.,  Houndjo, M. J. S., and  Tossa, J. (2019). 
New traversable wormhole solutions in $f(T)$ gravity. 
International Journal of  Modern Physics D, 28, 1950065. 


\bibitem{Yerlan} Myrzakulov, Y., Donmez, O., Yildiz, G. D. A., Güdekli, E., Muminov, S., and Rayimbaev, J. (2024). Linear redshift parametrization of deceleration parameter in $f (R, L_m)$ gravity. Physics of the Dark Universe, 45, 101545.
    \bibitem{Koussour} Koussour, M., Myrzakulov, N., Rayimbaev, J., Errehymy, A., and Donmez, O. (2024). Exploring accelerated expansion in the universe: A study of $f (Q, T)$ gravity with parameterized EoS and cosmological constraints. Chinese Journal of Physics, 90, 108-120.
\bibitem{Zhadyranova} Zhadyranova, A., Kanibekova, Z., Zhumabekova, V., Koussour, M., Anshokova, D., Muminov, S., and Rayimbaev, J. (2025). Observational evidence of bulk viscosity effects in $f (R, T)$ cosmological models. Physics Letters A, 548, 130560.



\end{thebibliography}
\end{document}